\documentclass[pra,aps,twocolumn]{revtex4}
\usepackage{amsmath,amssymb,amsthm,graphicx,hyperref}

\def\ket#1{\left| #1\right>}
\def\bra#1{\left< #1\right|}

\newtheorem{theorem}{Theorem}

\begin{document}

\title{Quantum States and Phases in Driven Open Quantum Systems with Cold Atoms}

\author{S. Diehl$^{1,2}$, A. Micheli$^{1,2}$,  A. Kantian$^{1,2}$, B. Kraus$^{1,2}$,  H.P. B\"uchler$^{3}$, and P. Zoller$^{1,2}$ }

\affiliation{$^1$Institute for Theoretical Physics, University of Innsbruck,
Technikerstr. 25, A-6020 Innsbruck, Austria}
\affiliation{$^2$Institute for Quantum Optics and Quantum Information of the
Austrian Academy of Sciences,  A-6020 Innsbruck, Austria}
\affiliation{$^3$Institute for Theoretical Physics III, University of Stuttgart,
Pfaffenwaldring 57, 70550 Stuttgart, Germany}

\date{\today}

\begin{abstract}
An open quantum system, whose time evolution is governed by
a master equation, can be driven into a given pure
quantum state by an appropriate design of the system-reservoir coupling.
This points out a route towards preparing many body states and non-equilibrium
quantum phases by quantum reservoir engineering. Here we discuss in
detail the example of a \emph{driven dissipative Bose Einstein
Condensate} of bosons and of paired fermions, where atoms
in an optical lattice are coupled to a bath of Bogoliubov excitations
via the atomic current representing \emph{local dissipation}.
In the absence of interactions the lattice gas is driven into a pure
state with long range order. Weak interactions lead to a weakly mixed
state, which in 3D can be understood as a depletion of the condensate,
and in 1D and 2D exhibits properties reminiscent of a Luttinger liquid
or a Kosterlitz-Thouless critical phase at finite temperature, with
the role of the ``finite temperature'' played by the interactions.
\end{abstract}

\maketitle

\section{Introduction}

In condensed matter physics quantum phases and associated strongly
correlated many body states are typically
prepared by cooling the system to low temperatures, 
where its properties are dominated by the ground state of a Hamiltonian, 
$H\ket{G}=E_{G}\ket{G}$,
i.e. by considering a \emph{thermodynamic equilibrium} situation,
where $\rho\sim e^{-H/k_{\mathrm{B}}T}\rightarrow\ket{G}\bra{G}$
for temperature $T\rightarrow0$. In particular, in the context of ultracold
atomic quantum gases, much of the interest of the last few years has focused on
\emph{engineering} \emph{specific Hamiltonians} based on control of microscopic
system parameters via external fields \cite{greiner02,stoferle04, kinoshita04,paredes04, folling07,zwierlein07}, opening the door to a
quantum simulation of strongly correlated ground
states \cite{jaksch98,petrov00,goral02,rezayi05,buechler05,rey07,bloch08}.

In contrast, quantum optics typically considers driven open quantum
system, where a system of interest is driven by an external field
and coupled to an environment inducing a \emph{non-equilibrium} dynamics,
with time evolution described by a master equation for the reduced
system density operator (see e.g. \cite{zollerbook}), 
\begin{eqnarray}
\dot{\rho} & = & -i[H,\rho]+{\cal L}\rho\label{eq:masterequation}\\ & \equiv &
-i[H,\rho]+\sum_{\ell}\kappa_{\ell}\left(2c_{\ell}\rho
c_{\ell}^{\dagger}-c_{\ell}^{\dagger}c_{\ell}\rho-\rho
c_{\ell}^{\dagger}c_{\ell}\right).\nonumber 
\end{eqnarray}
 Here $H$ is the
Hamiltonian of the driven system, while the Liouvillian
${\cal L}$ in Lindblad form represents the dissipative terms. The
\emph{quantum jump operators} $c_{\ell}$ are system operators as
they appear in the interaction Hamiltonian for the coupling to the
bath of harmonic oscillators, and describe the time evolution (quantum
jump) of the system associated with the emission of a quantum into
the harmonic oscillator bath with rate $\kappa_{\ell}$ in channel
$\ell$. The validity of the master equation is based on the Born-Markov
approximation with system-bath coupling in rotating wave approximation,
which in quantum optics is an excellent approximation because the
(optical) system frequencies are much larger than the decay rates.
For long times, the system described by (\ref{eq:masterequation})
will approach a dynamical steady state, $\rho(t)\rightarrow \rho_{ss}$,
which in general will be a mixed state. However, under special circumstances
the steady state can be a \emph{pure state,} $\rho_{ss}=\left|D\right>\left<D\right|$,
where in the language of quantum optics $\left|D\right>$ is called
a \emph{dark state}. Sufficient conditions for the existence of a
unique dark state are (i) $\forall\ell$ $c_{\ell}\left|D\right>=0$,
i.e. the dark state is an eigenstate of the set of quantum jump operators
with zero eigenvalue, which will be compatible with the dynamics induced
by the Hamiltonian if (ii) $H\left|D\right>=E\left|D\right>$. Uniqueness
of the dark state is guaranteed if there is no other subspace of the
system Hilbert space which is invariant under the action of the operators
$c_{\ell}$ \cite{baumgartner07,kraus08}. In fact, it can be shown that for any given pure state
there will be a master equation so that this state becomes the unique
steady state \cite{kraus08}. 

Here, we are interested in this novel possibility of quantum state engineering
by designing jump operators such that one drives the system into a desired
many-body quantum state. This {\it non-equilibrium} approach is in strong
contrast to conventional Hamiltonian engineering methods, as standard
thermodynamics concepts are not valid in this driven system, and the dynamics
goverened by the master equation (\ref{eq:masterequation}) is the only remaining
principle determining the final state. While in quantum optics we know several
examples of preparing \emph{single particle} pure states dissipation, including
dark state laser cooling to subrecoil temperatures \cite{aspect88,kasevich92}, 
it is of interest to extend these
ideas to many body systems, dissipatively driving the system into entangled
states of interest, or preparing non-equilibrium quantum phases in condensed
matter systems.  Furthermore, for the example of a dissipative driven Bose
Einstein condensate (BEC) discussed below, but also for stabilizer states in a
system of spins-$1/2$ or qubits living on a lattice \cite{kraus08}, the dissipation can be
chosen to be quasi--local, i.e. the jump operators act non--trivially only on a
small neighborhood of particles.

As an illustration of a many particle dark state, we discuss below a
dissipatively driven BEC, where for non-interacting atoms a pure state
exhibiting long range order is generated as the steady state by quasi-local
coupling to an environment with finite correlation length. Applying standard
linearization schemes in the weakly interacting situations, allows us to
determine the solution of the master equation (\ref{eq:masterequation}), and we
find that the steady state exhibits similar properties as bosons in thermal
contact to a heat bath: in 3D the effect of the interaction can be understood in
terms of a depletion of the condensate, while in 1D and 2D the system exhibits
properties reminiscent of a Luttinger liquid or a Kosterlitz-Thouless critical
phase at finite temperature.  In particular, we give a physical realization in
terms of cold bosonic atoms in an optical lattice by immersion in a superfluid
bath \cite{Griessner06}. As a second example we discuss a master equation whose
steady state corresponds to an $\eta$-condensate \cite{yang89,demler04}, i.e. a state of long range
order of paired fermions, which remarkably corresponds to an \emph{exact}
\emph{excited} eigenstate of a Hubbard Hamiltonian with repulsive or attractive
interactions in $d$ dimensions.

\section{long range order by local dissipation}

\subsection{Dissipative Driven Condensate}

\label{sec:DDBEC}Let us consider the dynamics of $N$ bosonic atoms
on $d$-dimensional lattice with $M^{d}$ lattice sites, lattice vectors
$\mathbf{e}_{\lambda}$ and spacing $a$ (see Fig. 1a). We assume
that the coherent motion of the atoms can be described a single band
Hubbard model with Hamiltonian\begin{equation}
H=H_{0}+V\equiv-J\sum_{\langle i,j\rangle}a_{i}^{\dagger}a_{j}+\frac{1}{2}U\sum_{i}a_{i}^{\dagger2}a_{i}^{2},\label{eq:Hubbard}\end{equation}
where $H_{0}$ represents the kinetic energy of bosons hopping between
adjacent lattice sites with amplitude $J$, and $V$ is the onsite
interactions with strength $U$, and $a_{i}$ ($a_{i}^{\dagger}$)
are bosonic destruction (creation) operators for atoms at site $i$.
A physical realization of this situation is achieved by loading cold
bosonic atoms into an optical lattice. By cooling to temperatures
$T\rightarrow0$ the system is prepared in the groundstate, which
for noninteracting atoms is $\left\vert \mathrm{BEC}\right\rangle =a_{\mathbf{q}=0}^{\dagger\, N}\left\vert \mathrm{vac}\right\rangle/\sqrt{N!}$
which corresponds to a state with macroscopic occupation of the quasimomentum
$\mathbf{q}=0$. Here $a_{\mathbf{q}}=\sum_{j}a_{j}e^{i\mathbf{q}
\mathbf{x}_{j}}/\sqrt{M^{d}}$
is the destruction operator for quasimomentum $\mathbf{q}$ in the
Bloch band. For weak interactions we have in 3D a BEC, and a quasi-condensate
in 1D and 2D, while with increasing interactions the system can undergo
a quantum phase transition to a Mott phase \cite{fisher89,jaksch98,greiner02}.

In contrast, we are interested here in a \emph{dissipative} Hubbard
dynamics modelled by a master equation (\ref{eq:masterequation}).
In atomic physics a bath will typically couple to the atoms via the
atomic density, $n_{i}=a_{i}^{\dagger}a_{i}$, as in the case of decoherence
due to spontaneous emission in an optical lattice, or for collisional
interactions. This will tend to dephase the condensate, and can heat
the system. In contrast, our goal is to couple the system to a bath
so that the system is driven to a pure many body state by quasi-local
dissipation. This is achieved, for example, by chosing jump operators
\begin{equation}
c_{\ell}\equiv c_{ij}=\left(a_{i}^{\dagger}+a_{j}^{\dagger}\right)\left(a_{i}-a_{j}\right)\label{eq:jump}
\end{equation}
 acting between a pair of adjacent lattice sites $\ell\equiv\langle i,j\rangle$
with a dissipative rate $\kappa_{ij}\equiv\kappa$ It is then easy
to see that the state $\left\vert \mathrm{BEC}\right\rangle $ satisfies
(i) $\forall\langle i,j\rangle$ $\ (a_{i}-a_{j})\left\vert \mathrm{BEC}\right\rangle =0$,
and (ii) is an eigenstate of the kinetic energy $H_{0}\left\vert \mathrm{BEC}\right\rangle =N\epsilon_{\mathbf{q}=0}\left\vert \mathrm{BEC}\right\rangle $,
where $\epsilon_{\mathbf{q}}=2J\sum_{\lambda}\sin^{2}\mathbf{q}\mathbf{e}_{\lambda}/2$
is the single particle Bloch energy for quasimomentum $\mathbf{q}$.
In other words, $\left\vert \mathrm{BEC}\right\rangle $ is a many
body dark state corresponding to a state of long range order for non-interacting
bosons on the lattice. In fact (see Appendix A), this state is
the unique steady state of this master equation, and any initial mixed
state will evolve for long times into $\left\vert \mathrm{BEC}\right\rangle $.

The key in obtaining a state of long range order as the steady state
is to couple to the bath involving the atomic current operator between
two adjacent lattice sites; the concept that dissipative coupling to the current operator
stabilizes superconductivity is well knwon in condensed matter \cite{schoen90}. The jump operator $c_{ij}$ describes
a pumping process where the second factor $a_{i}-a_{j}$ annihilates
the anti-symmetric (out-phase) superposition on the pair of sites
$\langle i,j\rangle$, while $a_{i}^{\dagger}+a_{j}^{\dagger}$ recycles
the atoms into the symmetric (in-phase) state. Loosely speaking, we
can interpret this process as a dissipative locking of the atomic
phases of two adjacent lattice sites, which in turn results in a global
phase locking, i.e. a condensate. Note that in view of $(a_{i}-a_{j})\left\vert \mathrm{BEC}\right\rangle =0$
it is the destruction part of the jump operator which makes $\left\vert \mathrm{BEC}\right\rangle $
the dark state, while any linear combination $a_{i}^{\dagger}$ and
$a_{j}^{\dagger}$ of recycling operators will do, except for a hermitian
$c_{ij}$ which would lead to a pure dephasing but no pumping into
the dark state.

The above discussion becomes particularly clear in a momentum representation
suggested by the translation invariance. The dissipative part of the
master equation takes on the form (\ref{eq:masterequation}) with
jump operators 
\begin{displaymath}
c_{\ell}\equiv c_{\mathbf{q},\lambda}=\frac{1}{\sqrt{M^{d}}}\sum\limits
_{\textbf{k}}(1+\mathrm{e}^{\mathrm{i}(\textbf{k} -
\mathbf{q})\textbf{e}_{\lambda}})
(1-\mathrm{e}^{-\mathrm{i}\textbf{k}\textbf{e}_{\lambda}}) 
a_{\textbf{k}-\mathbf{q}}^{\dagger}a_{\textbf{k}}
\end{displaymath}
and $\kappa_{\ell}\equiv\kappa$, which makes the appearance
of the dark state $\mathbf{q}=0$ decoupled from dissipation particulary
apparent. 

\subsection{Implementation}\label{sec:Implementation}

\begin{figure}
\begin{center}
\includegraphics[width=\columnwidth]{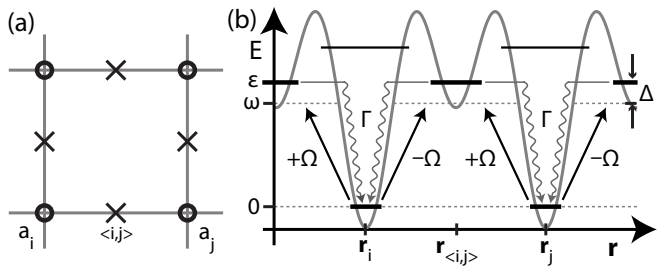}
\end{center}
\caption{\label{fig1} Driven Dissipative Condensate: (a) A lattice gas $a$ is immersed in a condensate $b$, which acts on the links $\langle i,j\rangle$ (crosses) of neighboring lattice sites $i$ and $j$ in the form a dissipative current. (b) Schematic realization of the effective dissipativ process in an optical super-lattice, which provides for excited states gapped by $\varepsilon$ and localized on the links  of neighboring lattice sites $\langle ij\rangle$: A Raman laser couples the  ground- and excited bands with effective Rabi-frequency $\Omega$ and detuning $\Delta=\omega-\varepsilon$ from the inter-band transition. The spatial modulation of the Raman-laser yields in a position dependent coupling, which excites only the anti-symetric component of atoms sitting on neighboring lattice sites $i$ and $j$ into the upper band. The inter-band decay with a rate $\Gamma$ back to their lower band is obtain via the emission of phonons into the surrounding BEC.}
\end{figure}

The above master equation can be realized by immersing atoms $a$
moving in an optical lattice in a large BEC of atoms $b$ \cite{Griessner06}. The condensate
interacts in the form of a contact potential with interspecies scattering length 
$a_{ab}$ with the atoms $a$, and acts as a bath of Bogoliubov excitations.
This coupling provides an efficient mechanism for decay of atoms $a$
from an excited to lower Bloch bands by emission of a Bogoliubov quasiparticle.
Atoms $a$ moving in the lowest Bloch band can be driven by laser
induced Raman processes to the first excited Bloch band, and can decay
back to the ground band. This situation is reminiscent of optical pumping in quantum
optics, or laser cooling \cite{aspect88,kasevich92,zollerbook}. There a laser excites
electronic states of an atom, which return to the ground state by
spontaneous emission of a photon. This formal analogy suggests and
justifies a description of a driven dissipative Hubbard dynamics in
terms of a master equation. Thus our goal is to identify an optical
lattice configuration which after adiabatic elimination of the excited
Bloch bands in the limit of weak driving results in a master equation
of the type discussed in Sec. \ref{sec:DDBEC}.

We consider a lattice as illustrated in Fig. 1a, with an additional
auxiliary lattice site on each of the links. The optical lattice corresponding
to a single link is shown in Fig. 1b. It has the form of a $\Lambda$-system
with the two Wannier functions of lattice sites 1 and 2 representing
two ground states, and the auxiliary state in the middle representing
an excited state. We drive this three-level system by Raman transitions
from the two ground to the excited states with Rabi frequencies $\Omega$
and $-\Omega$, respectively. The excited atom can decay back to 1
or 2 by emission of a Bogolibov quasiparticle. As is well-known from
quantum optics, such a $\Lambda$-configuration supports a dark state:
in the example of a single atom only the antisymmetric (out-of phase)
state $(a_{1}^{\dagger}-a_{2}^{\dagger})\ket{\textrm{vac}}$ is excited
by the laser, so that the atom is eventually ``pumped'' into the dark
symmetric (in-phase) state $(a_{1}^{\dagger}+a_{2}^{\dagger})\ket{\textrm{vac}}$.
In general, laser excitation followed by return of an atom to site
$\alpha=1,2$ will involve operators $a_{\alpha}^{\dagger}(a_{1}-a_{2})$,
and - as shown in Appendix B  - results in a Liouvillian with
the structure
\begin{eqnarray}\label{MasterConcrete}
{\cal L}\rho=\sum C_{\left\langle i,j\right\rangle ,\left\langle i',j'\right\rangle }^{\alpha\beta}\left[a_{\alpha}^{\dagger}(a_{i}-a_{j}),\rho(a_{i'}^{\dagger}-a_{j'}^{\dagger})a_{\beta}\right]+\textrm{{h.c.}}
\end{eqnarray}
The coefficients $C$ are related to the correlation function of the
Bogoliubov reservoir, and in particular exhibit the correlation of
emitted quasiparticles at lattice sites $\alpha$ and $\beta$ as
reflected by the correlation length of the reservoir. For wavelength
$\lambda_{\mathrm b}$ of Bogoliubov excitations larger or smaller than the
optical lattice spacing $a$, spontaneous emission is either correlated
or uncorrelated. However complicated $C$, the existence of a dark
state is guaranteed by $(a_{i}-a_{j})\left\vert \mathrm{BEC}\right\rangle =0$,
a property which follows from the laser excitation step. We will confine
our discussion below to the master equation with jump operators (\ref{eq:jump})
with qualitatively similar results expected for the general case.

\section{Competition of hamiltonian and Liouvillian Dynamics}
\label{sec:Competition}

For a realistic system with a finite interaction $V$, see Eq.~(\ref{eq:Hubbard}), the $|{\rm
BEC}\rangle$ is no longer a dark state of the master equation: the interactions
tend to localize the particles, while the dissipative terms tend to enforce a
pure condensate.  In general, the competition between these two incompatible
dynamics results in a mixed state.  Below we present linearized theories, which
allow us to solve for the density matrix $\rho$, 
as well as to study the correlations functions under the time evolution.

\subsection{Mean field theory}

For weak interactions, one can expect that the pure $|{\rm BEC}\rangle$ state is
only weakly perturbed with the zero momentum mode $a_{{\bf q = 0}}$ still
macroscopically occupied. We follow therefore the standard Bogoliubov prescription
and replace in the master equation the zero momentum mode by its mean value,
i.e., $a_{0}= \sqrt{n_{0} M^{d}}$ with $n_{0}$ the condensate density
($n-n_{0}\ll n$). In leading order, the jump operators in the master equation
(\ref{eq:masterequation}) reduce to
\begin{equation}
c_\ell = a_\ell, \hspace{20pt} c^{\dag}_\ell = a_\ell^{\dag},
\end{equation} 
and the coupling rates $\kappa_\ell \equiv    16 n \kappa \sum_{\lambda}
\sin^2\left({\bf q} {\bf e}_{\lambda}  /2\right)$. Here,  the index $\ell \!=\!\{{\bf
q}, \sigma\}$ with $\sigma\! =\! \pm 1$ characterizes the bosonic operators
$a_\ell\equiv a_{{\bf q},\sigma} = i^{1\!-\!\sigma} \left(a_{\bf q} \!+\!\sigma
a_{-{\bf q}} \right)/\sqrt{2}$.  In addition, the Hamiltonian in
Eq.~(\ref{eq:Hubbard})
simplifies to
\begin{equation}
 H = \sum_{\ell } \left\{\left( \epsilon_\ell+
Un \right) a^{\dag}_\ell a_\ell +\frac{U n}{2} \left[
a_\ell^{2}+\left(a^{\dag}_\ell\right)^{2} \right]
\right\}.
\label{boghamiltonian}
\end{equation}
with $\epsilon_\ell\equiv \epsilon_{\bf q}$. Note that $a_{{\bf q}, \sigma} =
\sigma a_{-{\bf q},\sigma}$ and the summation over the index $\ell$ avoids these
double countings.  It follows that the master equation decouples for each mode
$a_\ell$ and is quadratic in the bosonic operators. 
The analogous master equation in quantum
optics is well known from parametric amplification.  
The general solution is given by a mixed Gaussian state, which in
steady state takes the form $\rho = \mathcal{Z}^{-1} \Pi_\ell \rho_\ell$ with
\begin{eqnarray}
  \rho_\ell = \exp \left(- \beta_\ell b^{\dag}_\ell b_\ell \right)
\label{gaussiandensity}
\end{eqnarray}
with $\beta_\ell$ describing a finite occupation of the squeezed operators
$b_\ell =   e^{-i \phi_\ell} \cosh(\theta_\ell)\: a_\ell + e^{ i \phi_\ell}
\sinh(\theta_\ell)\:a^{\dag}_\ell$.
The squeezing parameter $\theta_\ell$ and $\beta_\ell$
are given by the relation
\begin{eqnarray}
 \cosh^2\left(2\theta_\ell\right)=
\coth^2\left(\beta_\ell/2\right)=\frac{\kappa_\ell^2+(\epsilon_\ell+ U
n)^2}{\kappa^2_\ell+E_\ell^2 }, 
\end{eqnarray}
with $E_\ell=\sqrt{\epsilon_\ell^2+ 2 U n \epsilon_\ell } \equiv E_{\textbf{q}}$ the Bogoliubov energy. The phase $\cot \phi_\ell = (\epsilon_\ell+n U)/\kappa_\ell$ plays a minor role. From these
relations, we recover the pure $|{\rm BEC}\rangle$ in the limit of vanishing
interactions $U n/\kappa
\rightarrow 0$. On the other hand for fixed interaction and dissipation $\kappa$,
the modes $b_\ell$ essentially reduce to the well known Bogoliubov modes for
small momentum $|{\bf q} \textbf{e}_{\lambda} |  \ll \sqrt{U J}/\kappa $, while the
parameter takes the form $\beta_\ell  \approx E_\ell/T_{\rm \scriptscriptstyle eff}$ with an effective temperature 
\begin{eqnarray}
T_{\rm \scriptscriptstyle eff} = U n/2.
\end{eqnarray}
The density matrix for the low momentum modes is therefore indistinguishable from
the thermal state of a weakly interacting Bose gas with the role of a ``finite
temperature'' played by the interactions $U n$.

This similarity of the driven system with a thermal Bogoliubov state naturally
rises the question on the validity of the mean field approximation: the ansatz assumes a small depletion $n_{\mathrm D}$ of the zero momentum
mode. Using the above solution for the steady state, the condensate depletion
takes the from
\begin{equation}
 n_{\mathrm D} = n-n_{0} = \frac{1}{2}\int \frac{d{\bf q} }{v_{0}} \frac{(U
n)^2}{\kappa_{\bf q}^2 + E_{{\bf q}}^2}
\end{equation}
with $v_{0}$ the volume of the Brillouin zone. This expression strongly depends
on the dimension of the system: in 3D the depletion remains finite and small for
weak interactions $U n /J\ll 1$. On the other hand, in one- and two-dimensions
an infrared divergence appears indicating the absence of a macroscopic
occupation and the breakdown of mean-field theory. Consequently, we find
that a dissipatively driven system exhibits the same behavior as a bosonic
system in themal contact with a heat bath, where the appearance of long-range
order at finite temperature is only possible above the lower critical dimension d=2.

The solution to the master equation also allows us to study the relaxation into
the steady state, see Appendix C. The build up of 
the macroscopic occupation  $n_{0}(t)$ obeys the long time behavior 
\begin{equation}
   n_{0}-n_{0}(t) \sim \sqrt{\frac{Un}{8 J}} \frac{1}{2 \kappa n t}.
\end{equation}
The condensate $n_{0}(t)$ approaches the steady state according to a
power law.  In the strict absence of interactions the behavior is modified to
$n_{0} - n_{0}(t) \sim t^{-3/2}$. The slower approach to equilibrium for the
interacting system results from the scrambling of the particles via the
interaction.

\subsection{Lower dimensions $d=1,2$}

In lower dimensions, the interaction drives strong phase fluctuations, which
destroy the macroscopic occupation of the condensate. However, these
fluctuations are only relevant on distances larger than the coherence length
$\sqrt{J/U n} a$ ($a$ denotes the lattice spacing). Consequently, the
formation of a local condensate still takes place on shorter distances, while
only the phase between these local condensates is destroyed by the fluctations.
The influence of these phase fluctuations can be studied in a long wave length
description by introducing a smoothly varying phase field $\phi_{i}$ and density
field $n_{i}$ with $\left[\phi_{i},n_{i}\right]= i \delta_{i j}$ \cite{haldane81}. This behavior
of the dissipatively driven system is in close analogy to the thermodynamics of
interacting bosons giving rise to Kosterlitz-Thouless critical phases in 2D, and
Luttinger liquids in 1D. The jump operators simplify to
\begin{equation}
c_{i j } = \left(n_{i}-n_{j}\right) - 2 i n \left(\phi_{i}-\phi_{j}\right),
\end{equation}
while the Hamiltonian (\ref{eq:Hubbard}) reduces to the harmonic model
\begin{equation}
 H =    J n \sum_{\langle i j \rangle}  \left(\phi_{i}- \phi_{j}\right)^2
 + \frac{U}{2} \sum_{i} n_{i}^2 .
\label{rotorhamilton}
\end{equation}
Consequently, the master equation becomes again quadratic and introducing new
bosonic operators $d_\ell$ with $\ell\! =\!\{{\bf q},\sigma\}$ allows us  to decouple
the master equation for each mode $\ell$: we define the bosonic operator $d_\ell=
i^{1-\sigma} \left(d_{\bf q}+\sigma d_{\bf -q}\right)$ with
%
 $d_{{\bf q}} = \left( n_{\bf q}/\sqrt{n}- i\sqrt{2 n}
\phi_{\bf q} \right)/\sqrt{2}$ .
%
Surprisingly, with this definition the jump operator reduces to $c_\ell \equiv
d_\ell$ and the master equation (\ref{eq:masterequation}) in the long wave length limit $|{\bf q}|a
<\sqrt{U n/J} $ is mapped to the same form as the master equation in the
previous section, with the operator $a_\ell$ replaced by the new operator $d_\ell$
and the chemical potential $U n$ replaced by $Un\! -\! \epsilon_\ell/2$.
Consequently, we find again the exact solution to the master equation, which
allows us to characterize the state via its correlation functions, see Appendix C; such a characterization of a states in cold gases has attracted a lot of interest
recently \cite{bistritzer07,burkov07,donner07,hofferberth07}.
 
First, we analyze the steady state. The interesting correlation function in
lower dimensions is the Green's function 
\begin{displaymath}
  G_{{\rm \scriptscriptstyle eq}}(x,t) = \langle a_{i}(t_{1})a_{j}^{\dag}(t_{0})\rangle \sim \langle
\exp\left[i \left( \phi_{i}(t_{1}) - \phi_{j}(t_{0})\right)\right]\rangle
\end{displaymath}
with  $x$  the distance between the lattice sites $i$ and $j$ and $t=|t_{1}-t_{0}|$. 
 In the long wave length limit with $x, \: tc  \gg \sqrt{J/U n} a$, the
smooth part of the correlation function recovers rotational and translation
invariance; here $c= \sqrt{2 U n J} a$ is the sound velocity. Then, 
simple analytic results are found in 2D (with the help of
the quantum regression theorem) giving rise to quasi long-range order
\begin{eqnarray}
G_{\rm \scriptscriptstyle eq}(x,0) & \sim & 
\left\{ 
\begin{array}{ccc}
(x_{0}/x)^{\frac{T_{\rm \scriptscriptstyle eff}}{4T_{\rm \scriptscriptstyle KT}}} & \hspace{10pt}&  x \gg c t\\
(\tau_{0}/t)^{\frac{T_{\rm \scriptscriptstyle eff}}{4T_{\rm \scriptscriptstyle KT}}} & \hspace{10pt}& x \ll ct 
\end{array}\right.\label{StStCorr}
\end{eqnarray}
with the Kosterlitz-Thouless temperature $T_{\rm \scriptscriptstyle KT}=\pi Jn\gg
T_{\rm \scriptscriptstyle eff}= U n/2$, and the short distances scales $x_{0} \sim c/ \kappa n$ 
and $\tau_{0} \sim (\kappa n a/c)^2 a/c$.  On the other hand, in one-dimension
we find an exponential decay both in space and time with coherence length $\xi_{{\rm
\scriptscriptstyle 1D}} = 4 c K/ \pi T_{\rm \scriptscriptstyle eff}$, with the parameter
$K=\pi\sqrt{2Jn/U}$ playing the role of the Luttinger parameter. The concept
of the effective temperature is particularly efficient in the low dimensional
systems, where the correlations are dominated by the low energy phase modes, as
implied by the exponential (1D) and algebraic decay (2D).  Noticeably,  the
functional dependence of the correlation function is determined by the ratio
$U/J$ alone, while the dissipative coupling strength $\kappa$ only gives rise to
non-universal prefactors. This behavior is a consequence of the low momentum
behavior of the dissipative damping $\kappa_{\textbf{q}} \sim {\bf q}^2$ versus the
linear sound spectrum of the Hamiltonian (\ref{rotorhamilton}).

\begin{figure}
\begin{center}
\includegraphics[width=\columnwidth]{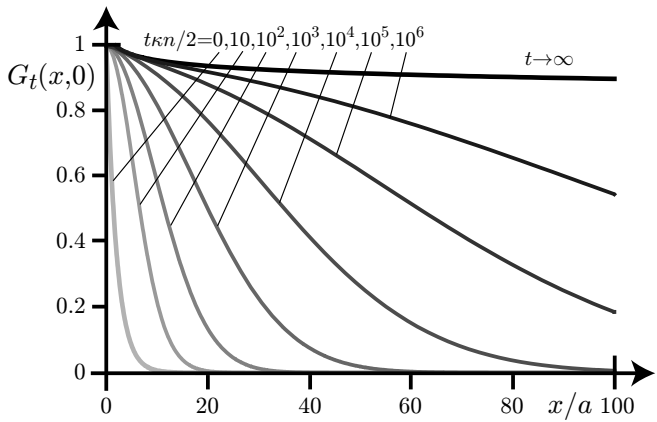}
\end{center}
\caption{\label{Fig:TimeEvol} Appearance of quasi long-range order during the time evolution:
the correlation function $G_t(x,0)$ is shown for various times $t \kappa n/2
= 0, 10,  10^2, 10^3,  10^4, 10^5, 10^6, \infty$. The intial disordered state has a correlation
length $\xi = 2 a$, and the system parameters are chosen at
$T_\mathrm{eff}/(4T_\mathrm{KT}) =1/18$ and
$x_0 = 0.55 a$.}
\end{figure}

Finally, we study the time evolution of the spatial correlation function in
two-dimensions i.e., $G_{t}(x,0) = \langle a_{i} a_{j}^{\dag}\rangle_{t}$
with the average defined by the density matrix $\rho(t)$.  The system is
initially prepared in a disordered state characterized by a small correlation
length $\xi$, giving rise to an exponential decay of the correlations.  The exact
time evolution then allows us to study the appearance of the quasi-long range
order Eq.~(\ref{StStCorr}) from this initially uncorrelated state.  For $t\gg
x^{2}/(16\kappa n)$, we obtain in leading order
\begin{equation}
G_{t}(x,0)  \sim (x/x_{0})^{-\frac{T_{\rm \scriptscriptstyle eff}}{4T_{\rm
\scriptscriptstyle KT}}} \: e^{-x^{2}/ x_{t}^2}.\label{Psi}
\end{equation}
The length scale within which the system  exhibits quasi-long range order is  $x_{t}=2(\pi\xi^{2}\kappa n a t)^{1/4}$ 
and increases in time with the universal power $x_{t} \sim t^{1/4}$, see Fig.~\ref{Fig:TimeEvol}.

\section{Condensate of interacting fermionic doublons}

As a second example, we consider the dissipative preparation of an
$\eta$-state, which is an {\em exact excited eigenstate} of the
$d$-dimensional two-species fermionic Hubbard-Hamiltonian  \cite{yang89}
\begin{equation}
H_{FH}=-J\sum_{\langle i, j\rangle
,\sigma}f_{i\sigma}^{\dagger}f_{j\sigma}+U\sum_{i}f_{i\uparrow}^{\dagger}f_{i\downarrow}^{\dagger}
f_{i\downarrow}f_{i\uparrow}.
\end{equation}
 This state is created
by the $\eta$-operator,
$\eta^{\dagger}=\sum_{i}\phi_{i}\eta_{i}^{\dagger}/M^{d/2}$
with the local doublon operators
$\eta_{i}^{\dagger}=f_{i\uparrow}^{\dagger}f_{i\downarrow}^{\dagger}$,
and $\phi_{i}=\pm 1$ denoting a sign alternating between sites in a
checkerboard pattern. The $N$-$\eta$-state is created by
$N$-fold application of $\eta^{\dagger}$, yielding an excited
eigenstate of $H_{FH}$, $\left(\eta^{\dagger}\right)^N|0\rangle$, with
energy $NU$. This excited state is a condensate of
doublons into the quasimomentum-state $(\pi,\ldots,\pi)$ in the corner
of the $d$-dimensional Brillouin-zone. It exhibits superfluidity, with
non-decaying
off-diagonal long-range order in any spatial dimension. Together with the operator $\sum_{\sigma} f^{\dagger}_{i\sigma}f_{i\sigma}$, the
properties
of $\eta^{\dagger}$ allow the construction of symmetry
generators of $H_{FH}$, which has been used to investigate high-$T_c$
superconductivity
(c.f. \cite{demler04}).

Following the lines presented above, we construct quasi-local
operators at each pair of lattice sites having the $N$-$\eta$- state
as a dark-state. As is easily verified, the quasilocal operators
$c_{\ell}\equiv c_{ij}^{(k)}$ ($k=1,2$) given by
\begin{eqnarray}
c_{ij}^{(1)}& = &(\eta_{i}^{\dagger}-\eta_{j}^{\dagger})(\eta_{i}+\eta_{j}),\\
c_{ij}^{(2)}& = &
n_{i\uparrow}f_{i\downarrow}^{\dagger}f_{j\downarrow}+n_{j\uparrow}f_{j\downarrow}^{\dagger}f_{i\downarrow},
\end{eqnarray}
fulfill these requirements. Small-scale numerical simulations for
open-boundary systems then show that a Liouvillian (\ref{eq:masterequation})
constructed from these quantum jump-operators
and the Hamiltonian $H_{FH}$ drive any initial
$\rho(0)$ into the $N$-$\eta$-state, assuming that $N_{\uparrow}=N_{\downarrow}$
at all times. The result may be interpreted in the quantum
jump picture: $H_{FH}$ generates configurations with spin-up and
spin-down particles on adjacent sites from any initial configuration.
These configurations are then captured by $c_{ij}^{(2)}$, associating
them into doublons. Subsequent action of $c_{ij}^{(1)}$
then generates the desired $\eta$-state by phase-locking,
just as in the case of the jump-operators for the BEC.

\section{Conclusions and Outlook}

\label{sec:Conclusion}

We have discussed a scenario where many body quantum states and entangled
states are prepared as dark states in a non-equilibrium driven dissipative
dynamics with quasi-lcoal dissipation. While the present work has
focused on condensed matter aspects of realizing non-equlibirium (quasi-)condensates
of interacting bosons and paired fermions in optical lattices, the
present ideas are readily extended to spin systems, and promise a
new avenue towards preparing interesting entangled states of qubits
for quantum information \cite{kraus08}. On the atomic physics side,
control via external fields offers interesting new possibilities of
engineering a broad class of quantum jump operators (\ref{eq:jump}),
one example being phase imprinting $a_{i}\rightarrow a_{i}e^{i\phi_{i}}$
with a laser. 

\emph{Acknowledgements} We thank E. Altman, E. Demler and M. Lukin for discussions. Work at the University of Innsbruck is supported by the Austrian Science Foundation and EU grants SCALA and OLAQI.

\appendix
\label{sec:Methods}

\section{Criterion for the Uniqueness of the Stationary State (see Sec. \ref{sec:DDBEC}) [16]}

The BEC is a dark state for the set of jump operators $c_\ell$. For the uniqueness of this solution we further need it to be the only dark state.  For a fixed particle number $N$, the first factor in $c_\ell$ (see Sec. (\ref{sec:DDBEC})) has no eigenvalues and in particular, no zero eigenvalues. Thus, in order to identify dark states $| D\rangle$ with zero eigenvalue, we may restrict to the equation $(a_i - a_j)|D\rangle = 0 \,\,\forall\langle i,j \rangle$. Taking the Fourier transform, this translates to $(1 - e^{i{\bf q} \textbf{e}_\lambda a})a_{\bf q}  |D\rangle = 0 \,\,\forall\textbf{q}$. Thus the BEC state with $\textbf{q} =0$ is the only dark state. 

For the dark states to be the only stationary solution of $\dot{\rho} =\mathcal{L}\rho$ we need to show now that if there exists another
stationary state, then there must exist a subspace of the Hilbert
space ${\cal H}$, which is invariant under any application of the
operators $c_\ell$. Equivalently, in \cite{kraus08} we show the following

\begin{theorem} Let $D$ be the space of dark states, i.e.  $c_\ell D =0$. If there exists no subspace $S\subseteq{\cal H}$
with $S\perp D$ such that $c_\ell S\subseteq S$ $\forall c_\ell$,
then the only stationary states are the dark states.\end{theorem}

In order to show that the jump operators Eq. (\ref{Jump!}) indeed lead to the unique BEC steady state solution we construct a polynomial operator $O(c_{\ell})$ with $\bra{BEC}O\Phi\rangle\neq0$ for all $\ket{\Phi}\in S=\mathcal{H}\backslash\ket{BEC}$. For this purpose we work in the momentum space representation  such that 
\begin{eqnarray*}\label{MomRep}
c_{\mathbf{q},\lambda}=\frac{1}{\sqrt{M^{d}}}\sum\limits _{\textbf{k}}(1+\mathrm{e}^{\mathrm{i}(\textbf{k}-\mathbf{q})\textbf{e}_{\lambda}a})(1-\mathrm{e}^{-\mathrm{i}\textbf{k}\textbf{e}_{\lambda}a})a_{\textbf{k}-\mathbf{q}}^{\dagger}a_{\textbf{k}}.
\end{eqnarray*}
with $\textbf{e}_\lambda$ characterizing the different lattice vectors connecting nearest neighbour sites. A basis in the Hilbert space can be written as $\ket{\{n_{\mathbf{q}}\}}=\prod_{\mathbf{q}}(a_{\mathbf{q}}^{})^{n_{\mathbf{q}}}\ket{0}$.
Therefore the polynomial operator $O=\prod_{\mathbf{q}}c_{\mathbf{q}}^{n_{\mathbf{q}}}$
provides a finite overlap with the BEC. A general state can be written
as $\ket{\Phi}=\sum f_{\{n_{\mathbf{q}}\}}\ket{\{n_{\mathbf{q}}\}}$.
Choose one basis state with nonvanishing coefficient with highest
occupation in the zero mode. Then the polynomial operator $O$ for
this basis state provides also a finite overlap of $\ket{\Phi}$ with
the dark state.

\section{Implementation of the BEC Liouvillian}
\label{methods:Implementation}

{\em Coherently driven two-band Hubbard model} -- We outline the mechanism for engineering Liouvillians driving into a BEC. We focus on a one-dimensional system, where the system atoms $a$ are moving in an optical superlattice $V_{\rm opt}=\sum_{n=1}^2V_n\sin^2(n\pi z/a)/n^2$ 
with lattice spacing $a$ along the direction $z$, while being tightly confined by a harmonic potential with oscillator frequency $\omega_\perp$ in the transverse directions, $x$ and $y$. The lattice depths $V_2>V_1>0$ of the superlattice are chosen such that the vibrational spacings $\hbar\omega_{n=0,1}\approx2\sqrt{E_{\rm r}(V_2{\pm}V_1)}$ about the individual wells at positions $z_{0,j}=ja$ and $z_{1,j}=(j+1/2)a$ are much larger than their gap $\varepsilon \approx V_1+\hbar(\omega_1-\omega_0)/2$, cf. Fig.\ref{fig1}(b). Here $E_{\rm r}=\hbar^2\pi^2/2m_aa^2$ denotes the lattice recoil and $m_a$ the mass of the atoms $a$.  The low-energy dynamics of the atoms $a$ is then given by a two-band Bose-Hubbard model 
\begin{align}
H_a=&\sum_{n,q}\bar\varepsilon_{n,q} \bar a_{n,q}^\dag \bar a_{n,q}+\tfrac{1}{2}\sum_{\{\alpha_i\}} U_{\{\alpha_i\}} a_{\alpha_1}^\dag a_{\alpha_2}^\dag a_{\alpha_3}^\dag a_{\alpha_4}\nonumber
\end{align}
where $\bar a_{n,q}^\dag=\sum_{j=1}^M e^{i q j a}a_{n,j}^\dag/\sqrt{M}$ creates a Bloch-wave with quasi-momentum $0<q\leq\pi/a$ and in the lower and upper band,  $n=0,1$, respectively, and $a^\dag_{n,j}$ creates an atom $a$ in the site at position $z_{n,j}$. We denote the dispersion relations for the two bands by $\bar\varepsilon_{n,q}=-2J_n\cos(qa)+n\varepsilon$ with $J_n$ the nearest neighbor inter-band hopping rates in band $n$ and $\varepsilon$ the band-separation. 
The second term in $H_a$ with $\alpha_i\equiv(n_i,j_i)$ account for intra- and inter-band interaction of two atoms $a$ in the superlattice in terms of the respective on- and off-site shift $U_{\{\alpha_i\}}$. The dominant intra- and intra-band interactions are given by density-density interactions $\sim\sum_{n,j} U_n a_{n,j}^{\dag 2}a_{n,j}^2/2$ and $\sim V_{0,1}\sum_{(ij)} a_{0,j}^\dag a_{1,j}^\dag a_{1,j} a_{0,j}$, respectively. Here, $U_n$ denotes state-dependent on-site shifts, and $V_n$ the intra-band off-site shift for nearest neighbors $(ij)$, c.f. $i-j=0,1$.

A key ingredient for the BEC Liouvillian is a selective coherent drive between the two bands of the Hubbard model, which couples the antisymmetric superposition on a pair of lower sites to the upper level on the link in between. This is achieved via a Raman laser setup in the form of a set of standing waves locked at the position of one of its minima $z_{1,j}$. This allows to realize an effective dynamical coupling $V_{\rm las}(t)=\hbar\Omega\cos(\pi z/a)\sin(\omega t)$ with the same periodicity as the optical lattice. Here $\Omega$ is the two-photon Rabi frequencies, and we denote the Raman-detuning from the upper band by $\Delta=\omega -\varepsilon/\hbar$. For a weak driving field, $\Omega \ll \varepsilon/\hbar$, the Raman-drive  results in effective inter-band coupling 
\begin{align}
H_{\rm las}(t) =& \hbar \bar\Omega e^{-i\omega t} \sum_j (-1)^j  a_{1,j}^\dag \left(a_{0,j} - a_{0,j+1}\right) + {\rm H.c.}\nonumber\\
=&\hbar\bar\Omega e^{-i\omega t} \sum_q (1-e^{iqa})\bar a^\dag_{1,q+\pi/a}\bar a_{0,q} + {\rm H.c.}\nonumber
\end{align}
where we made use of the rotating wave approximation and dropped AC-Stark shifts, since they only renormalize the gap $\sim V_1$ of the original superlattice, and $\bar\Omega=\Omega f$ is the effective Rabi-frequency up to a Franck-Condon factor $f$, which in terms of the Wannier-functions $w_n(z-z_{n,j})$ for the band $n$ localized at $z=z_{n,j}$ reads $f=\int dz w_0(z) \cos(\pi z/a) w_1(z-a/2)$. The alternating signs in the interband-couping $H_{\rm las}$ for atoms sitting on neighboring sites in the lower band $n=0$ originates from the modulation of the Raman-coupling. This results in a excitation of asymmetric (and in particular antisymetric) superposition of atoms sitting on adjacent sites from the lower band to the upper band, while their symmetric superposition is dark with respect to $H_a$, leading to the BEC dark state. From  momentum representation of $H_{\rm las}(t)$  we see that the two counter-propagating components $e^{\pm i\pi z/a}$ provide a inter-band coupling of modes differing by $\pi/a$, and their interference results in the Bloch-waves to be selectively excited from the lower to the upper band based on their quasi-momentum $q$. In particular for the lower-band $n=0$ the two components destructively (constructively) interfere for $q=0$, ($q=\pi/a$), while the upper-band displays the opposite behavior.

\emph{Coupling to the phonon reservoir} -- The dissipative step is implemented by immersing the system into a large homogeneous 3D condensate of a distinguishable species of atoms $b$, which acts as a reservoir of phonon modes \cite{Griessner06}. The corresponding phonon Hamiltonian is given in the Bogoliubov approximation by $H_b = \sum_{{\bf k}\neq0} E_k b^\dag_{\bf k} b_{\bf k}$, where $b_{\bf k}$ creates a Bogoliubov excitation with momentum ${\bf k}$ and energy $E_k=[(\hbar kc_b)^2+(\hbar^2k^2/2m_b)^2]^{1/2}$ with $m_b$ the mass of the atoms $b$ and $c_b$ their speed of sound in the condensate. The atoms $a$ and $b$ interact via a contact-interaction  with a coupling constant $g_{ab}=2\pi\hbar^2a_{ab}/\mu_{ab}$ given in terms of their intra-species scattering length $a_{ab}$ and their reduced mass $\mu_{ab}=m_am_b/(m_a+m_b)$. Expanding these density-density interaction in terms of the fluctuations about the condensate wave-function we obtain to first order (apart from an overall state-independent mean-field shift $g_{ab}N_a\rho_b$ proportional to the condensate density $\rho_b$) an effective coupling of the atoms $a$ to the Bogoliubov excitations in the form
\[ H_{ab} = g_{ab}\sum_{{\bf k}\neq0} \sqrt{\rho_b S_k} A_{\bf k}^\dag b_{\bf k} +{\rm H.c.}\]
Here $S_k=\hbar^2k^2/2m_bE_k$ is the static structure factor of the BEC and $A_{\bf k}^\dag$ is the displacement-operator with momentum ${\bf k}$ for the atoms $a$ associated with the recoil from the emission of a Bogoliubov excitation into the condensate.

In the following we are interested in the effective dynamics of the system atoms $a$ and consider the BEC as a reservoir of Bogoliubov excitation at essentially zero temperature, since under typical experimental conditions one can achieve temperatures $T_{\rm b}\ll\varepsilon/k_{\rm b}$. We integrate out the bath dynamics in the Born-Markov approximation, and obtain a Master-equation for the density operator $\rho(t)$ of the atoms $a$ (within the rotating wave and independent rate approximation with respect to the laser excitation),
\begin{align}
\frac{d\rho}{dt}=& \frac{1}{i\hbar}[H_a(t),\rho] + \frac{1}{V}\sum_{\bf k} |\frac{g_k}{\hbar}|^2 ([A_{\bf k},\rho \bar A^\dag_{\bf k}(E_k)]+{\rm H.c.}),\nonumber\\
\bar A_{\bf k}^\dag(\varepsilon)&= \int_0^\infty d\tau e^{+i(\varepsilon-H_a)\tau/\hbar}A_{\bf k}^\dag e^{+iH_a\tau/\hbar}\nonumber
\end{align}
with $H_a(t)=H_a+H_{\rm las}(t)$ and where $A_{\bf k}^\dag(\varepsilon)$ denotes the Fourier-component of $A_{\bf k}$ with frequency $\varepsilon/\hbar$ with the frequency $E_{k}/\hbar$ given by the dispersion relation of the Bogoliubov excitations. Given that intra-band dissipative processes are suppressed by momentum conservation \cite{Griessner06}, we focus on the inter-band decay, and within the rotating frame approximation write the Master-equation with $A_{\bf k}^\dag=\sum_q G_{{\bf k},q}^{(1,0)}\bar a^\dag_{1,q+k_z}\bar a^\dag_{0,q}$ and their Fourier-components given by $A_{\bf k}^\dag(E)\approx A_{\bf k}^\dag \left[\pi\delta(E/\hbar-\varepsilon/\hbar)+i{\cal P}/(E/\hbar-\varepsilon/\hbar)\right]$, where we neglected corrections $J_\sigma,U_\sigma\ll \varepsilon$ in the spectra and in the following will drop the Lamb-shifts, as they amount to (small) second-order shifts.

We take the continuum limit of ${\bf k}$ and exploiting the radial symmetry of the Bogoliubov spectrum $E_k$, perform the integration over ${\bf k}$ are left with a integral over its azimutal angle, which we rewrite in terms of $k_z/k$ as
\begin{align}
{\cal L}\rho 
=& \sum_{q,q'} \frac{\pi k_0^2}{v_0}\int d^2n  G^{(1,0)*}_{k_0 {\bf n},q} G^{(1,0)}_{k_0 {\bf n},q'}\times\nonumber\\
&\left[ \bar a_{0,q}^\dag \bar a_{1,q+k_0n_z}, \rho \bar a_{1,q'+k_0n_z}^\dag \bar a_{0,q'}\right]+{\rm H.c.},\nonumber
\end{align}
assuming a tight transverse confinement, $\varepsilon\ll\hbar\omega_\perp=\hbar^2/2m_aa_\perp^2$, we approximate the Bloch-function by a Gaussian of width $a_\perp$ in transverse direction times periodic component along $z$, $\Phi_{\sigma,q}({\bf r})\approx \phi_{\sigma,q}(z) e^{-\rho^2/2a_\perp^2}/(\pi a_\perp^2)^{1/4}$ and perform the polar integration in $G^{(1,0)}$ as
\begin{align}
G^{(1,0)}_{{\bf k},q} &\approx g_ke^{-(k^2-k_z^2) a_\perp^2/4}f^{(1,0)}_{k_z,q}/\hbar,\nonumber\\
f^{(1,0)}_{k_z,q}&= \int dz \phi_{1,q+k_z}^*(z)e^{ik_zz}\phi_{0,q}(z)\nonumber
\end{align}

{\em Long wavelength (Super/Sub-radiant) limit} --
In general the recoil along $z$ for the bloch-wavefunction has to be computed numerically. However in the deeply bound limit a rough estimate may obtained by taking (orthogonalized) gaussians of width $a_z$ for the wannier functions and focus on nearest neighbor contributions, which yields
\begin{align}
f^{(1,0)}_{k,q}\approx& \sum_{l=0,1} e^{-i q l a} \int dz e^{ikz} w_1(z)w_0(z-(2l-1)a/2) \nonumber\\
\approx &\frac{2\exp[-(ka_z/2)^2]\sin^2(ka/8)e^{-iqa/2}}{\sinh[(a/4a_z)^2]/\cos(ka/4-qa/2)}.\nonumber
\end{align}

Thus the master-equation in momentum space reads
\begin{align}
{\cal L}\rho =& \kappa_0\int_{-k_0}^{+k_0} dk_z  e^{-k_z^2(a_z^2-a_\perp^2)/2}\sin^4(\tfrac{k_za}{8})\times\nonumber\\
&\sum_{q,q'}e^{i(q-q')a/2}\cos\left(\tfrac{k_za}{4}-\tfrac{qa}{2}\right)\cos\left(\tfrac{k_za}{4}-\tfrac{q'a}{2}\right)\times\nonumber\\
&\left[ \bar a_{0,q}^\dag \bar a_{1,q+k_z}, \rho \bar a_{1,q'+k_z}^\dag \bar a_{0,q'}\right]+{\rm H.c.},\nonumber
\end{align}
with $\kappa_0=2\pi k_0 g_{k_0}^2e^{-k_0^2a_\perp^2/2}/\hbar^2v_0\sinh^2[(a/4a_z)^2]$. We notice that for $k_0<\pi/a$, one obtains a collective intra-band decay decribed by a set of jump operators $C_k$ with momentum transfer $|k|\leq k_0<\pi/a$ along the lattice  as 
\begin{eqnarray}
C_k = \sqrt{\kappa_k} \sum_q e^{i q a/2}\cos\left(\tfrac{ka}{4}-\tfrac{qa}{2}\right)\bar a_{0,q}^\dag \bar a_{1,q+k},\nonumber
\end{eqnarray} 
with $\kappa_k=\kappa_0e^{-k^2(a_z^2-a_\perp^2)/4}\sin^2(ka/8)$.

{\em Short wavelength limit} -- For $k_0>n\pi/2$ the sum over $k$ spans over $n$ Brillouin zones, resulting in construct and destructive interference, which one associates to the finite correlation length of the Bogoliubov modes in the bath. In particular, for $k_1\gg k_0=\pi/a$ we notice that the summation in the master-equation runs over several Brillouin zone, which effectively suppresses the non-local / long-range decay. Thus it is convenient to rewrite the master-equation in position space, which yields the decay of the general form
\begin{align}
{\cal L}\rho = \sum_{i,j}\sum_{\ell,\ell'} \Gamma_{j,j'}^{\ell,\ell'} \left[a_{0,j}^\dag a_{1,\ell}, \rho a_{1,\ell'}^\dag a_{0,j'}\right] + {\rm H.c}\nonumber
\end{align}
where the summations $j,j'$ and $\ell,\ell'$ are over the lattice sites and links, respectively, and in the limit $k\rightarrow\infty$ the decay rates take the simple form 
\begin{align}
\Gamma_{j,j'}^{\ell,\ell'}=\Gamma_0 \int dz w_{0,j}w_{1,\ell}w_{1,\ell'}w_{0,j'},\nonumber
\end{align} 
where  $\Gamma_0 =g_{ab}S_{k_1}/1-S_{k_1}^2$ in terms of the  being the static structure factor $S_{k_1}$ at momentum $k_1$ and $w_{n,i}\equiv w_n(z-z_{n,i})$ denote the Wannier function for the lower ($n=0$) and upper ($n=1$) band, respectively.
We notice that they are translationally invariant and rapidly fall off with increasing separation of $(j\ell\ell'j')$. Thus we can restrict ourself to the largest quasi-local contributions, an in particular the ones for neighboring lattice sites yield 
\begin{align}
{\cal L} \rho = \sum_j \sum_{p=0,1} &\left( \gamma  \left[a_{0,j}^\dag a_{1,j+p},\rho a_{1,j+p}^\dag a_{0,j}\right]+\right.\nonumber\\
&\left. \gamma_0 \left[a_{0,j+p}^\dag a_{1,j},\rho a_{1,j}^\dag a_{0,j+1-p}\right] +\right.\nonumber\\
&\left.\gamma_1 \left[a_{0,j}^\dag a_{1,j-p},\rho a_{1,j-1-p}^\dag a_{0,j}\right]+{\rm H.c.}\right).\nonumber
\end{align}
The first term describes the uncorrelated decay from the link $j$ to the left site $j$ and to the right site $j+1$ with $\gamma=\Gamma_{j,j}^{j,j}=\Gamma_{j+1,j+1}^{j,j}$. The second one accounts for the finite correlation between the two processes (cf. $\sim J_0$) and the third is an analogous one that  accounts for correlations in the decay between different links to the same site, i.e. links $j-1$ and $j$ to site $j$  (cf. $\sim J_1$).
It is convenient to rewrite the three-terms in terms of symmetric / antisymmetric combinations of particle localized on adjacent sites in the two bands, cf. $(a_{n,j\mp1}\pm a_{n,j+1\mp1})/\sqrt{2}$. This results in a master-equation in the form of a ``chain'' of concatenated $\Lambda$-systems and inverted $\Lambda$ systems,
\begin{align}
&{\cal L} \rho = %
\sum_{j,\pm} \left(\frac{\gamma}{2}\pm\gamma_0\right)%
\left[\frac{a_{0,j}\pm a_{0,j+1}}{\sqrt{2}}a_{1,j},\rho%
a_{1,j}^\dag\frac{a_{0,j}\pm a_{0,j+1}}{\sqrt{2}}\right]  \nonumber\\
&+%
\sum_{j,\pm} \left(\frac{\gamma}{2}\pm\gamma_1\right)%
\left[a_{0,j}\frac{a_{1,j-1}\pm a_{1,j}}{\sqrt{2}},\rho%
a_{1,j}^\dag\frac{a_{1,j-1}\pm a_{1,j}}{\sqrt{2}}a_{0,j}\right]\nonumber\\
&+ {\rm H.c.} \nonumber
\end{align}
corresponding to a master-equation with jump operators $C^{\Lambda\langle ij \rangle}_{\pm}$ on links ($\Lambda$-systems) and vertices (V-systems) respectively, as
\begin{subequations}
\begin{align}
C^{\Lambda_{\langle ij \rangle}}_{\pm} =& \frac{\sqrt{\gamma\pm2\gamma_0}}{2}(a_{0,j}^\dag \pm a_{0,j+1}^\dag) a_{1,j} \nonumber\\
C^{{\rm V}_{\langle j\rangle}}_\pm =& \frac{\sqrt{\gamma\pm2\gamma_1}}{2}a_{0,j}^\dag  (a_{1,j-1} \pm a_{1,j})\nonumber
\end{align}
\end{subequations}

In the independent rate approximation we thus obtain the master-equation for the set of laser-driven excitations and quasi-local decay (that is in the regime where the wavelength of the Bogoliubov excitations is much smaller than the lattice-spacing, $k_1\gg \pi/a$)
by 
\begin{subequations}
\begin{align}
\frac{d\rho}{dt} =& \frac{1}{i\hbar}\left[H_a + H_{\rm las}(t),\rho(t)\right]+{\cal L}_\Lambda \rho + {\cal L}_V\rho,\nonumber\\
{\cal L}_\alpha \rho=& \sum_{\langle i,j \rangle^\alpha_\pm} \left[ 2 C^{\alpha_{ij}}_\pm\rho (C^{\alpha_{ij}}_\pm)^\dag - \{ (C^{\alpha_{ij}}_\pm)^\dag  C^{\alpha_{ij}}_\pm,\rho\} \right],\nonumber
\end{align}
\end{subequations}
where $\alpha_{ij}$ denote the summation over links and (connected) vertices for $\alpha=\Lambda$ and $\alpha=V$, respectively.

From the master-equation we see that (i) all excited states decay back to the ground-state and (ii) for $U=0$ the phase-locked ground state (BEC) is a dark-state, i.e. an eigenstate of the hamiltonian that is dark with respect to the excitation and the decay.

Since all excited states are decaying, we adiabatically eliminate the excited band. In the limit that one has a weak far detuned laser, $\Delta\gg\Omega,U_n,J_n,\gamma_\pm^\Lambda,\gamma_\pm^{\rm V}$, results to lowest order
$a_{1,j}\approx \Omega (a_{0,j}-a_{0,j+1})/\sqrt{2}\Delta$ and thus
the jump-opeators transform to (up to global phases)
\begin{align}
C_{\lambda(j,\pm)} =& \sqrt{\kappa^\Lambda_\pm} (a_{0,j}^\dag \pm a_{0,j+1}^\dag)(a_{0,j}-a_{0,j+1})/2,\nonumber\\
C_{{\rm v}(j,\pm)} =& \sqrt{\kappa^{\rm V}_\pm} a_{0,j}^\dag( -a_{0,j-1}+(1\pm1)a_{0,j} \mp a_{1,j+1})/2,\nonumber
\end{align}
where we dropped the band-index $n=0$ and we introduced the rates $\kappa_\pm^\alpha=\gamma_\pm^\alpha(\Omega/\Delta)^2$, and remark that the latter correspond to the diagonalized form of Eq.~\eqref{MasterConcrete}.

\section{Momentum Space Correlations}

\label{HarmApprox}

In the linearized theory all information is encoded in the first and second moments. Our construction
implies the vanishing of the first moments, $\langle a_\ell^{\dagger}\rangle=\langle a_\ell\rangle=0$.
We may therefore concentrate on the time evolution equations of the
second moments, whose solution then allows to reconstruct the Gaussian
density operator and gives access to the correlation functions. In
principle the relevant single particle sector of the density matrix
is mapped out by the correlations $\langle a_\ell^{\dagger}a_{\ell'}\rangle,\langle a_\ell a_{\ell'}\rangle,h.c.$,
but for our purposes it is sufficient to focus on the diagonal entries.
The corresponding equations are obtained from the linearized version of Eq. (\ref{eq:masterequation}) (cf. Sect. \ref{sec:Competition}) using the commutation relation $[a_\ell ,a_{\ell'}^\dagger] = \delta_{\sigma,\sigma'}(\delta_{\textbf{q},\textbf{q}'} + \sigma\delta_{\textbf{q},-\textbf{q}'} ), [a_\ell ,a_{\ell'}]  =[a_\ell^\dagger ,a_{\ell'}^\dagger]  =0$ ($\ell = (\textbf{q},\sigma)$), with the result
 \begin{eqnarray}
\partial_{t}\langle a_\ell^{\dagger}a_{\ell}\rangle & = & 2\, \mathrm i Un(\langle a_\ell a_\ell\rangle - \langle a_\ell^{\dagger}a_\ell^{\dagger}\rangle ) - 4\kappa_\ell   \langle a_\ell^{\dagger}a_\ell\rangle ,\label{TimeEvol}\\\nonumber
\partial_{t}\langle a_\ell a_{\ell}\rangle & = & - 4\mathrm i [Un(\langle a_\ell^\dagger a_\ell\rangle +\tfrac{1}{2}) + (\epsilon_\ell + Un)\langle a_\ell a_\ell\rangle ]\\\nonumber 
&& - 4\kappa_\ell   \langle a_\ell a_\ell\rangle 
 \end{eqnarray}
where we omit the equation for $\langle a_\ell^{\dagger}a_\ell^{\dagger}\rangle$ trivially obtained form the second line. The equations of motion may be solved via the Laplace transform and form the basis for the computation of the density operator and the spatial and temporal correlation functions.

\end{document}